\documentclass[twocolumn,showpacs,preprintnumbers,amsmath,amssymb]{revtex4}

\usepackage{graphicx}
\usepackage{dcolumn}
\usepackage{bm}

\begin{document}


\title{Multiphoton discrimination at telecom wavelength with charge integration photon detector }

\author{Mikio Fujiwara}%
\email{fujiwara@nict.go.jp}
\author{Masahide Sasaki}

\affiliation{%
National Institute of Information and Communications Technology
Koganei, Tokyo 184-8795, Japan
}%

\date{\today}

\begin{abstract}
We present a charge integration photon detector (CIPD) that enables the efficient measurement of photon number states at the telecom-fiber wavelengths with a quantum efficiency of 80\% and a resolution less than 0.5 electrons at 1 Hz sampling. The CIPD consists of an InGaAs PIN photodiode and a GaAs JFET in a charge integration amplifier, which is cooled to 4.2 K to reduce thermal noise and leakage current. The charge integration amplifier exhibits a low noise level of 470 $nV/Hz^{1/2}$. The dark count is as low as 500 electrons/hour.
\end{abstract}

\maketitle


A photodetector that can count the number of photons in a pulse, a so-called photon number resolving detector (PNRD) is an essential tool for all quantum optics experiments and quantum info-commnunication technologies. Recent theoretical developments have shown that single photons, combined with linear optics and PNRDs, can be used to implement scalable quantum computers\cite{A1}. It has also been shown that PNRDs are needed to establish a universal gate set to synthesize any desired optical non-linearity by combining squeezed light beams, linear and quadratic optical interactions, homodyne detectors, and feedforward controls\cite{A2, A3}. Such non-linear gates are essential to achieve quantum collective decoding\cite{A2}, which is able to boost the capacity of a communications channel\cite{A4} as well as enhance the practical security level of a quantum cryptographic system. Unfortunately, PNRD technology is not yet mature. Efforts to achieve PNRD using several device types and materials continue. 
In the visible region, a visible light photon counter (VLPC)\cite{A5} can count the number of photons despite a large dark count. Waks and his coworkers succeeded in observing non-classical photon statistics in parametric down conversion by VLPC\cite{A6}. A device based on time-resolving counters with fiber storage loops, beam splitters and commercially available Si avalanche photo diodes (APDs) has been demonstrated\cite{A7} as another practical scheme. The main obstacle in this scheme is to reduce various losses in the fibers and the couplers. 
In the telecom-fiber wavelength band, quantum key distribution\cite{A8, A9} and quantum teleportation\cite{A10} have been demonstrated so far, an InGaAs APD single photon detector with gated Geiger mode is usually employed. Its quantum efficiency is, however, about 10\%, and the counting rate is limited to about 1 MHz due to the after pulse phenomenon\cite{A11}. The InGaAs ADP single photon detector is thus under development.
 A superconducting transition-edge sensor was developed as a PNRD\cite{A12}. While it has sensitivity for a wide range of wavelengths and low dark counts, it requires special process techniques and has to be cooled to 100 mK. Moreover, the quantum efficiency is ~20\%, which is not high enough for measuring the statistics of non-classical light directly. 
In this letter, we describe a charge integration photon detector (CIPD) that enables the efficient measurement of photon number states at telecomm-fiber wavelengths. Our system is constructed with an InGaAs PIN photodiode (Kyosemi Corp., Japan) and a GaAs junction field effect transistor (JFET) (SONY Ltd., Japan) as a preamplifier for the charge integration circuit. The schematic is shown in Fig. 1. A framed area in this figure is cooled to 4.2 K to reduce noise and leakage current. A PIN photodiode converts one photon into one electron-hole pair with a high regard for linearity to incident photons. Therefore, if a low noise amplifier is employed, the number of photons can be accurately estimated by counting the number of charges. The gate electrode of the JFET is connected to the InGaAs PIN photodiode, so that the GaAs JFET operates as a source follower. Photo-carriers are integrated in the gate electrode, and the gate voltage and the source voltage increase at a rate proportional to the number of integrated carriers. Integration amplifiers generally need reset FETs to release charges. Such devices, however, increase the input capacitance and noise, which degrade the signal to noise ratio as described later. In our scheme, therefore, we adopt a mechanical reset probe. That is, the reset probe is isolated from the circuit, and touches the electrode only at the moment of discharge. 

\begin{figure}
\includegraphics{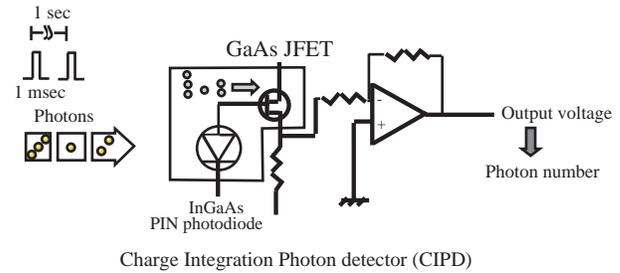}
\caption{\label{fig:epsart} Conceptual view of charge integration photon detector (CIPD).}
\end{figure}

An InGaAs PIN photodiode has responsivity with high quantum efficiency (~80\%) for the wavelengths from 900 nm to 1700 nm at room temperature. The spectral response of the PIN photodiode at three kinds of temperatures is shown in Fig. 2(a). The cutoff wavelength grew shorter as the operating temperature was reduced, and reached 1540 nm at 4.2 K. This may be because phonon assisted photon absorption is reduced, while the band gap in InGaAs increases as the operating temperature decreases. However, the responsivity at 1530 nm is slightly higher at 4.2 K than that at room temperature. Comparing with data sheet provided by the maker, the quantum efficiency at 4.2 K at 1530 nm is as high as 80\%. Unfortunately, the responsivity at 1550 nm was not high enough, at which the attenuation rate of an optical fiber is minimum. Band engineering is needed to increase the quantum efficiency of the InGaAs PIN photodiode for the whole telecom-fiber waveband at cryogenic temperatures. 

The signal to the noise ratio in this system is given as follows,
\begin{align}
Singal&=GM \cdot NQ/C_{input} \notag \\
Noise&=V_{noise}	\\
S/N&=GM \cdot NQ/(C_{input}V_{noise}) \notag
\end{align}

where \textit{GM} is the source follower gain, Cinput is the input capacitance, \textit{Q} is the elemental electric charge, \textit{N} is the number of carriers, and Vnoise is the channel noise of the GaAs JFET. To improve the resolution, decreasing the input capacitance and the channel noise is essential. The capacitance of the InGaAs PIN photodiode as a function of inverse bias voltage at room temperature and 4.2 K are shown in Fig. 2(b). The diameter of the PIN photodiode is 30 $\mu m$. The capacitance decreased to as low as 0.024 pF at 4.2 K. In addition, the gate capacitance of the GaAs FET whose gate width and length are 5 $\mu m$ and 50 $\mu m$, respectively, is 0.043 pF. We used a hollow mounting technique, and attained a total input capacitance as small as 0.067 pF. 

\begin{figure}
\includegraphics{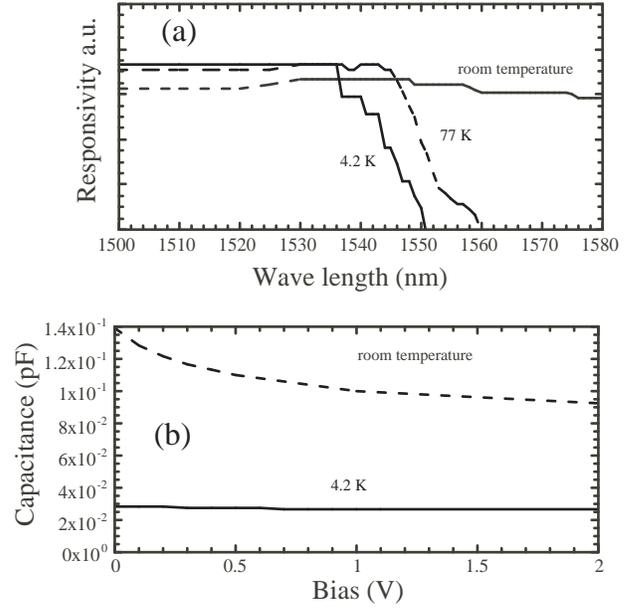}
\caption{\label{fig:epsart}   (a) Spectral response of  InGaAs  PIN photodiode with operation temperature as a parameter. 
(b) Capacitance of  InGaAs PIN photodiode as a function of reverse bias voltage at room temperature and 4.2 K.}
\end{figure}

In order to reduce the channel noise, we applied the thermal cure technique\cite{A13}, which can effectively suppress the random telegraph signal noise of the GaAs JFET. The noise level we obtained was ~470 $nV/Hz^{1/2}$ at 1 Hz. The noise spectrum of the CIPD is shown in Fig. 3(a). The noise level at the sampling rate can be estimated from this spectrum. For an integration readout circuit, correlated double sampling (CDS)\cite{A14} is used for estimating the  noise voltage as well as the output voltage. The input charges are estimated by comparing the initial output voltage with that after an integration time T. When the noise power in CDS is calculated, two kinds of filter effects must be considered. First, sampling procedure is expressed as $g(t)=\delta (t+T)-\delta (t)$, where the Fourier transform is given by
\begin{align}
|F(f)|&=\Bigl| \int_{-\infty}^{\infty}g(t)exp(-2\pi ft)dt \Bigr|, \notag \\
&=2|sin(\pi ft)|
\end{align}
which means that CDS acts as a comb filter. Moreover, the output voltage is averaged over a period $T_0$. $T_0$ can be as long as $T-\Delta$ where $\Delta$ is a pulse width (irradiated time). This procedure also has a filter effect, and it is expressed as
\begin{align}
h(t)&= \begin{cases}
		1\quad -T_0/2<t<T_0/2  \notag \\
		0\quad t<-T_0/2,\quad t>T_0/2 \notag \\
		\end{cases} \notag \\
|H(f)|&=\Bigl| \int_{-\infty}^{\infty}h(t)exp(-2\pi ft)dt \Bigr|,  \\
&=\Biggl|\frac{sin(\pi T_0f)}{\pi f} \Biggr| \notag
\end{align}

Combining $F(t)$ and $H(t)$, and the noise voltage at $CDS_{Vnoise, CDS}$ is given by

\begin{equation}
V_{noise,CDS}^2=\int_0^{\infty}\frac{V_{noise}^2 \bigl|F(f)^2H(f)^2\bigr|}{1+\left(\frac{f}{f_c} \right)^2}df \quad,
\end{equation}

where $f_c$ is the cutoff frequency of the low pass filter in the circuit. In our experiment, $T=1 \mathrm{s}$, $\Delta =10 \mathrm{ms}$, and  $f_c =20 \mathrm{Hz}$. The noise voltage is evaluated as 0.8 mV when the power spectrum near 0 Hz has \textit{1/f} dependency. Compared with this value, a single photo-carrier generates 2 mV, taking into account the \textit{GM} of 0.85 and the total input capacitance of 0.067 pF. Hence, our system can discriminate the photo-carrier number with a resolution of 0.5 electrons. The direct measurement of the standard deviation of the output voltage in the dark condition at a sampling of 800 events results in 0.24 electron/s, whose distribution is shown in the inset of Fig. 3(a). The dark count of the CIPD was 500 electrons/hour.

\begin{figure}
\includegraphics{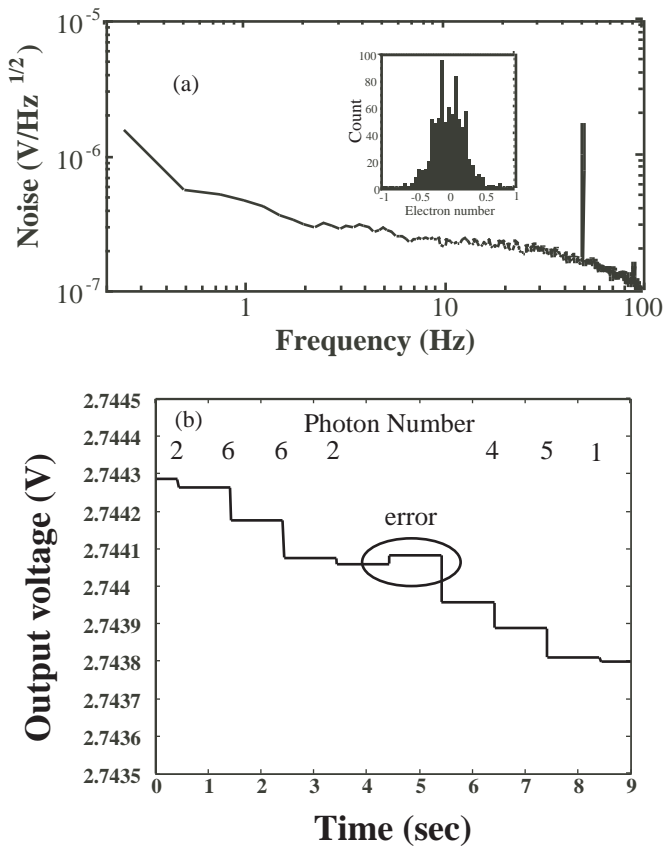}
\caption{\label{fig:epsart}   (a) Noise spectrum of CIPD system. (b) Time variation of average output voltage signal. Inset shows the deviation of the dark counts.}
\end{figure}

For testing photon counting, we injected heavily attenuated coherent light pulses with an interval of 1 s into the detector and recorded the output voltage. Incident light from a CW laser was shaped into a pulse with a 10 ms width by a lithium niobate EO modulator. The average number of photons in the pulse was adjusted to a few ~ a few tens. The signal light is incident on the InGaAs PIN photodiode in the cryostat through a single mode fiber with a focuser. The coupling efficiency is up to 80\%. The time variation of the average output voltage signal is shown in Fig. 3(b). The voltage step heights correspond to the number of the input photo-carriers, which are shown in the upper inset. The step with the increasing voltage, marked by the oval, is an error due to the random telegraph signal noise. 
Histograms of the measured photon number (precisely, the photo-carrier number) distribution for five kinds of pulse intensities are shown in Fig. 4. 

\begin{figure}
\includegraphics{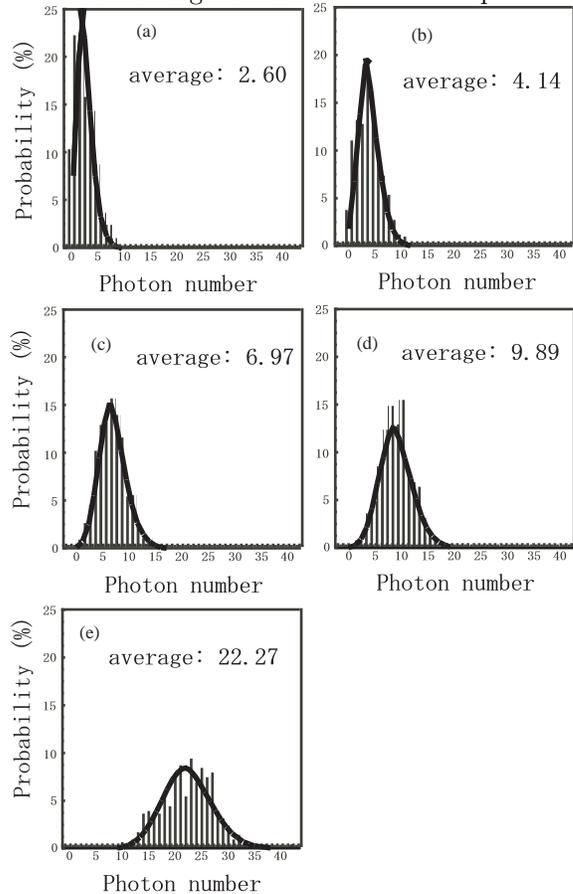}
\caption{\label{fig:epsart}   Photon number distribution: average photon number and sampling number are (a) 2.60, 744; (b) 4.14, and 569; (c) 6.97, 796; (d) 9.89, 800; (e) 22.27, 399.}
\end{figure}
The sampling numbers ranged from about 400 to 800. The solid lines indicate the theoretical fitting with Poisson statistics. The close match between the measured data and the fittings means that we can reproduce the expected dependence of Poisson statistics on pulse intensity. The quantum efficiency of the InGaAs photodiode obtained by back calculation is $80\pm 5\%$. 
In conclusion, we have developed a charge integration photon detector (CIPD) consisting of a InGaAs PIN photodiode and a GaAs JFET that can count the number of photons in a telecom-fiber wavelength with quantum efficiency of 80\% and a resolution less than 0.5 electrons at 1 Hz sampling.

\end{document}